\documentclass[prd,aps,preprint,tightenlines,nofootinbib,superscriptaddress]{revtex4}

\usepackage{graphicx}
\usepackage{dcolumn}
\usepackage{bm}

\begin {document}

\title{The elliptic gluon GTMD inside a large nucleus}

\author{Jian~Zhou}
\affiliation{\normalsize\it School of physics, $\&$ Key Laboratory of Particle Physics and Particle
Irradiation (MOE), Shandong University, Jinan, Shandong 250100, China}

\begin{abstract}
\noindent We evaluate the elliptic gluon Generalized Transverse Momentum Dependent(GTMD)
distribution inside a large nucleus using the McLerran-Venugopalan model. We further show that this
gluon distribution can be probed through the angular correlation in virtual photon quasi-elastic
scattering on a nucleus.
\end{abstract}

\maketitle

\section{introduction}
The quantum phase space distribution of partons inside a nucleon plays a central role in exploring
the tomography picture of nucleon. For a fast moving nucleon, a five-dimensional phase space Wigner
distribution which carries the complete information on how a single parton is distributed inside a
 nucleon has been introduced in the
literatures~\cite{Ji:2003ak,Belitsky:2003nz,Lorce:2011kd,Lorce:2011ni}. The Fourier transform of
the Wigner distributions  referred to as the Generalized Transverse Momentum Dependent(GTMD)
distributions~\cite{Meissner:2009ww} are  normally considered as the mother distributions of
Transverse Momentum Dependent(TMD) distributions and Generalized Parton Distributions(GPDs). So
far, the studies of GTMDs mostly focus on the formal theory side, including model
calculations~\cite{Meissner:2009ww,Courtoy:2013oaa,Kanazawa:2014nha,Mukherjee:2014nya,Liu:2015eqa,Mukherjee:2015aja,Chakrabarti:2016yuw},
the analysis of the multipole structure associated with GTMDs~\cite{Lorce:2013pza,Lorce:2015sqe} as
well as the investigation of their QCD evolution properties~\cite{Echevarria:2016mrc}. Perhaps most
interestingly, it has been revealed that one of GTMDs denoted as $F_{1,4}$~\cite{Meissner:2009ww}
can be related to the parton canonical orbital angular momentum~\cite{Lorce:2011kd,Hatta:2011ku}.

Recently, the issue how to access GTMDs experimentally is attracting  growing  attentions. In the
context of  small $x$ formalism, the impact parameter dependent unintegrated gluon distributions
often show up in the cross sections of diffractive
processes~\cite{Brodsky:1994kf,Buchmuller:1998jv,Martin:1999wb,Kovchegov:1999kx,Kowalski:2003hm,Marquet:2007qa,Lappi:2010dd,Rezaeian:2012ji}.
 The equivalence  of gluon TMDs and small $x$
unintegrated gluon distributions has been first established in Ref.~\cite{Dominguez:2011wm}, and
further clarified in Ref.\cite{Zhou:2016tfe}.  Following the similar procedure, one could also
identify the impact parameter dependent unintegrated gluon distributions as gluon Wigner
distributions due to the same operator structure~\cite{Hatta:2016dxp}. Therefore, one is allowed to
probe gluon GTMDs in various high energy diffractive scattering processes once a small $x$
factorization framework is employed.

Since the impact parameter $b_\perp$ and the gluon transverse momentum $q_\perp$ are left
unintegrated in the current case, one can define a new gluon Wigner distribution, the so-called
elliptic gluon Wigner distribution~\cite{Hatta:2016dxp} associated with the nontrivial angular
correlation $2(q_\perp \cdot b_\perp)^2-1$. The integrated version of this gluon distribution is
known as the helicity flip gluon GPDs~\cite{Hoodbhoy:1998vm,Belitsky:2000jk}. It has been shown
that the elliptic gluon distribution naturally emerges after implementing  the impact parameter
dependent BFKL/BK
evolution~\cite{Lipatov:1985uk,Navelet:1997tx,Navelet:1997xn,GolecBiernat:2003ym,Marquet:2005zf,Kopeliovich:2007fv,Hatta:2016dxp}(for
a previous detailed numerical analysis, see Ref.~\cite{Hagiwara:2016kam}). This finding inspires us
to construct a saturation model for the elliptic gluon distribution which can be used as a proper
initial condition for the small $x$ evolution. In addition to this motivation, a semi-classical
model calculation of the elliptic gluon distribution would be helpful for deepening our
understanding how the tomography picture of nucleon/nucleus is affected by the saturation effect.
To be more specific, we will compute the elliptic gluon GTMD in the McLerran-Venugopalan(MV)
model~\cite{McLerran:1993ka} that has been widely used to calculate the both unpolarized and
polarized small $x$ gluon
TMDs~\cite{Kovchegov:1996ty,JalilianMarian:1996xn,Metz:2011wb,Dumitru:2015gaa,Zhou:2013gsa}. In the
end, we point out that this gluon distribution is accessible through a $\cos 2\phi$ azimuthal
asymmetry in virtual photon-nucleus quasi-elastic scattering. Such measurement can be performed at
the future EIC.

The rest of the paper is structured as follows. In the next Section, we present some details of the
evaluation of the gluon elliptic GTMD in the MV model. In Sec.~III, we derive the azimuthal
dependent cross section for the virtual photon-nucleus quasi-elastic scattering.  The paper is
summarized in Sec.~IV.

\section{The elliptical gluon GTMD in the MV model }
As well known, gluon TMDs are process dependent and correspondingly possess the different gauge
link structure. The same statement applies  to gluon GTMDs as well. In the present work, we focus
on discussing  the dipole type gluon GTMD, which is defined as the
following~\cite{Mulders:2000sh,Hatta:2016dxp},
 \begin{eqnarray}
xG_{DP}(x,q_\perp, \Delta_\perp)&=&2 \int \frac{d \xi^- d^2 \xi_\perp e^{-iq_\perp\cdot \xi_\perp
-ixP^+\xi^-}}{(2\pi)^3 P^+} \nonumber \\ && \times  {\biggl \langle} P+\frac{\Delta_\perp}{2}
{\biggl |} {\rm Tr} \left [ F^{+i}(\xi/2) U^{[-]\dag} F^{+i}(-\xi/2) U^{[+]} \right ]
{\biggl|}P-\frac{\Delta_\perp}{2} {\biggl \rangle}
\end{eqnarray}
where $\Delta_\perp$ is the transverse momentum transfer to nucleus.  Here the longitudinal
momentum transfer to nucleus is ignored.  The gauge link in the current case takes a closed loop
form  in the fundamental representation.  In the small $x$ limit, up to the leading logarithm
accuracy the above expression can be reduced to~\cite{Hatta:2016dxp},
\begin{eqnarray}
xG_{DP}(x,q_\perp, \Delta_\perp)=\left ( q_\perp^2-\frac{\Delta_\perp^2}{4} \right )
\frac{2N_c}{\alpha_s} \Phi_{DP}(q_\perp, \Delta_\perp)
\end{eqnarray}
with $\Phi_{DP}(q_\perp, \Delta_\perp)$ being given by,
\begin{eqnarray}
\Phi_{DP}(x,q_\perp, \Delta_\perp)=\int \frac{d^2 b_\perp d^2 r_\perp}{(2\pi)^4} e^{-iq_\perp \cdot
r_\perp-i\Delta_\perp \cdot b_\perp} \frac{1}{N_c}{\biggl \langle}  {\rm Tr} \left [
U(b_\perp+\frac{r_\perp}{2}) U^\dag(b_\perp-\frac{r_\perp}{2}) \right ]{\biggl \rangle}
\end{eqnarray}
From a phenomenological point of view, the correlation limit where $ |\Delta_\perp|  \ll |q_\perp|$
is the most interesting kinematical region.  In a such kinematical limit,  depending on the angular
correlation structure, one can parameterize $\Phi_{DP}(x,q_\perp, \Delta_\perp)$ as,
\begin{eqnarray}
\Phi_{DP}(x,q_\perp, \Delta_\perp)={\cal F}_x(q_\perp^2, \Delta_\perp^2)+  \frac{q_\perp \cdot
\Delta_\perp}{|q_\perp||\Delta_\perp|}O_x(q_\perp^2,\Delta_\perp^2)+ \left [\frac{(q_\perp \cdot
\Delta_\perp)^2}{q_\perp^2 \Delta_\perp^2} -\frac{1}{2}\right ] {\cal F}^{\cal E}_x(q_\perp^2,
\Delta_\perp^2)+...
\end{eqnarray}
The Fourier transform of the first term on the right side of the equation $\int d^2 \Delta_\perp
e^{i\Delta_\perp \cdot b_\perp} {\cal F}_x(q_\perp^2, \Delta_\perp^2)$ is the normal impact
parameter dependent dipole type unintegrated gluon distribution. The second term
$O_x(q_\perp^2,\Delta_\perp^2)$ is a T-odd(or C-odd) distribution and commonly referred to as the
odderon exchange. The expectation value of the spin independent odderon has been computed in a
quasi-classical model and shown to be proportional to the slope of the saturation
scale~\cite{Kovchegov:2012ga}. For a transversely polarized target, one can introduce a spin
dependent odderon associated with the angular correlation $q_\perp \cdot S_\perp$ where $S_\perp$
is the target transverse spin vector~\cite{Zhou:2013gsa,Szymanowski:2016mbq}. It has been shown in
Refs.\cite{Boer:2015pni,Boer:2016xqr} that three different T-odd gluon TMDs can be related to the
spin dependent odderon. We refer readers  to Ref.~\cite{Boer:2016fqd} for the relevant
phenomenology studies of these distributions. The third term, namely the elliptic gluon GTMD ${\cal
F}^{\cal E}_x(q_\perp^2,\Delta_\perp^2)$ is what we are interested to compute in the MV model.
Higher order harmonics that could exist are not shown in the above equation. Note that the BFKL
dynamics only produces even harmonics, while one may expect that odd harmonics could be generated
by taking into account the BKP evolution~\cite{Bartels:1980pe} which describes the asymptotical
behavior of the C-odd gluon exchange at high energy.

To gain some intuition how the elliptic gluon distribution emerges from a quasi-classical
treatment, it would be instructive to first compute it in the dilute limit. Expanding the Wilson
line to the first non-trivial order(neglecting the tadpole type diagram for the moment), one has,
\begin{eqnarray}
\Phi_{DP}(x,q_\perp, \Delta_\perp)= \!\!\! \int \frac{d\xi_1^- d\xi_2^- d^2 b_\perp d^2
r_\perp}{(2\pi)^4} e^{-iq_\perp \cdot r_\perp-i\Delta_\perp \cdot b_\perp} \frac{g_s^2}{2N_c}
{\bigr \langle} A^+_a(\xi^-_1, b_\perp+\frac{r_\perp}{2}) A^+_a(\xi_2^-,b_\perp-\frac{r_\perp}{2})
{\bigr \rangle}
\end{eqnarray}
Following the standard procedure, we solve the classical Yang-Mills equation and express the gauge
field in terms of the color source,
\begin{eqnarray}
A^+_a(\xi_1^-,b_\perp+\frac{r_\perp}{2})=\int d^2 y_\perp G(b_\perp+\frac{r_\perp}{2}-y_\perp)
\rho_a(\xi_1^-,y_\perp)
\end{eqnarray}
with
\begin{eqnarray}
 G(k_\perp)=\int  d^2 \xi_\perp e^{-i\xi_\perp \cdot k_\perp} G(\xi_\perp)=\frac{1}{k_\perp^2}
\end{eqnarray}
We proceed to evaluate the color source distribution by a Gaussian weight function $ W_A[\rho]=
\exp \left [-\frac{1}{2} \int d^3 x \frac{ \rho_a(x) \rho_a(x)}{\lambda_A(x^-)} \right ] $. This
leads to,
\begin{eqnarray}
\Phi_{DP}(x,q_\perp, \Delta_\perp) \!\! &=& \!\! g_s^2C_F \!\! \int \frac{ d^2 b_\perp d^2 r_\perp
d^2 x_\perp}{(2\pi)^4} e^{-iq_\perp \! \cdot r_\perp-i\Delta_\perp \! \cdot b_\perp}
 G(b_\perp\! +\! \frac{r_\perp}{2}\! -x_\perp) G(b_\perp\!-\! \frac{r_\perp}{2}\!-x_\perp)
\mu_A(x_\perp)\nonumber \\ &=&\frac{g_s^2 C_F}{(2\pi)^4 }
\frac{1}{(q_\perp-\Delta_\perp/2)^2(q_\perp+\Delta_\perp/2)^2} \mu_A(\Delta_\perp)
\end{eqnarray}
where $  \mu_A(\Delta_\perp)=\int d^2 x_\perp e^{i\Delta_\perp \cdot x_\perp} \mu_A(x_\perp)$ with
 $\mu_A(x_\perp)=\int d\xi_1^- \lambda_A(\xi_1^-,x_\perp)$.
In the correlation limit, we can Taylor expand the above formula in terms of the power
$|\Delta_\perp|/|q_\perp|$. According to the parametrization for $\Phi_{DP}(x,q_\perp,
\Delta_\perp)$, it is easy to find the following expressions for the gluon GTMDs in the dilute
limit,
\begin{eqnarray}
{\cal F}_x(q_\perp^2, \Delta_\perp^2) &=& \frac{\alpha_sC_F}{4\pi^3}
\frac{\mu_A(\Delta_\perp)}{q_\perp^4}
 \nonumber \\  {\cal F}^{\cal E}_x(q_\perp^2, \Delta_\perp^2)&=&
\frac{\Delta_\perp^2}{q_\perp^2} {\cal F}_x(q_\perp^2, \Delta_\perp^2) \label{dilute}
\end{eqnarray}
In the below, we will use the above results as the base line to compare with the results in the
saturation regime.

To extend the analysis to the saturation regime, we have to take into account all intial/final
state interactions encoded in the Wilson lines.  We still follow the standard procedure to evaluate
the Wilson lines in the MV model. The contributions from the tadpole type diagram should be
included in the current case. The dipole amplitude in the MV model then reads,
\begin{eqnarray}
\Phi_{DP}(x,q_\perp, \Delta_\perp)&=& \!\! \int \frac{d^2 b_\perp d^2 r_\perp}{(2\pi)^4}
e^{-iq_\perp\cdot r_\perp-i\Delta_\perp \cdot b_\perp}
 \nonumber\\&\times& \!\!
 e^{ -g_s^2 C_F \!\! \int d^2x_\perp\mu_A(x_\perp)
 \left [\frac{G(b_\perp+\frac{r_\perp}{2}-x_\perp)^2}{2}
 +\frac{G(b_\perp-\frac{r_\perp}{2}-x_\perp)^2}{2} -
 G(b_\perp+\frac{r_\perp}{2}-x_\perp) G(b_\perp-\frac{r_\perp}{2}-x_\perp)\right ] }
 \nonumber\\
 &=&\int \frac{d^2 b_\perp d^2 r_\perp}{(2\pi)^4} e^{-iq_\perp\cdot
r_\perp-i\Delta_\perp \cdot b_\perp}
 e^{ -\alpha_s  C_F \frac{1}{8\pi} \int d^2x_\perp\mu_A(x_\perp)
  \left [ {\rm ln} \frac{(b_\perp+\frac{r_\perp}{2}-x_\perp)^2}{(b_\perp-\frac{r_\perp}{2}-x_\perp)^2}
 \right ]^2 }
\end{eqnarray}
The above expression can be further simplified by making the following approximations valid in the
correlation limit. In the perturbative region, the fact that $|r_\perp| \ll |b_\perp| \sim
|x_\perp|$ allows us to Taylor expand the logarithm term,
\begin{eqnarray}
&&\int d^2x_\perp\mu_A(x_\perp)
  \left [ {\rm ln} \frac{(b_\perp+\frac{r_\perp}{2}-x_\perp)^2}{(b_\perp-\frac{r_\perp}{2}-x_\perp)^2}
 \right ]^2 \nonumber \\&& \approx
 \int_{r_\perp}^{1/\Lambda_{QCD}} d^2 y_\perp
 \left\{ \frac{4 r_\perp^2 \cos^2(\phi) }{y_\perp^2}
 +\frac{2[4\cos^4(\phi)-3\cos^2(\phi)] r_\perp^4}{3y_\perp^4} \right \}\mu_A(b_\perp-y_\perp)
 \label{a}
\end{eqnarray}
where $\phi$ is the azimuthal angle between $r_\perp$ and $y_\perp \equiv b_\perp-x_\perp $. Since
the integration is dominated by  small $y_\perp$ region,  one can make a Taylor expansion in terms
of $y_\perp$,
\begin{eqnarray}
\mu_A(b_\perp-y_\perp)\approx \mu_A(b_\perp)-\frac{\partial \mu_A(b_\perp)}{\partial b_\perp^i}
y_\perp^i+\frac{1}{2} \frac{\partial^2 \mu_A(b_\perp)}{\partial b_\perp^i\partial b_\perp^j}
y_\perp^i y_\perp^j
\end{eqnarray}
Inserting it into the Eq.\ref{a}, we obtain,
\begin{eqnarray}
&&\int d^2x_\perp\mu_A(x_\perp)
  \left [ {\rm ln} \frac{(b_\perp+\frac{r_\perp}{2}-x_\perp)^2}{(b_\perp-\frac{r_\perp}{2}-x_\perp)^2}
 \right ]^2 \nonumber \\ & \approx &
2\pi r_\perp^2 {\rm ln} \frac{1}{r_\perp^2 \Lambda_{QCD}^2} \mu_A(b_\perp) \nonumber
\\ &&+ \frac{\pi  r_\perp^2}{24} {\rm ln} \frac{1}{r_\perp^2 \Lambda_{QCD}^2}
\left \{ 2 \left(  r_\perp \cdot \frac{\partial}{\partial b_\perp} \right
)^2 - r_\perp^2 \frac{\partial}{\partial b_\perp} \cdot  \frac{\partial}{\partial b_\perp} \right
\} \mu_A(b_\perp)
\end{eqnarray}
The simple power counting tells us that the second term scales as $\Delta_\perp^2Q_s^2/q_\perp^4$.
In the kinematical region where $Q_s|\Delta_\perp| \ll q_\perp^2$, it is not necessary to sum the
second term into the exponential form. Based on this observation, the dipole distribution can be
expressed as,
\begin{eqnarray}
\Phi_{DP}(x,q_\perp, \Delta_\perp)&=& \!\! \int \frac{d^2 b_\perp d^2 r_\perp}{(2\pi)^4}
e^{-iq_\perp\cdot r_\perp-i\Delta_\perp \cdot b_\perp} \nonumber \\ &\times & \!\! \left [1
-\frac{r_\perp^2}{192} \left \{ 2 \left(  r_\perp \cdot \frac{\partial}{\partial b_\perp} \right
)^2 - r_\perp^2 \frac{\partial}{\partial b_\perp} \cdot  \frac{\partial}{\partial b_\perp} \right
\} Q_s^2(b_\perp) \right ]
 e^{ -\frac{r_\perp^2 Q_s^2(b_\perp)}{4}}
\end{eqnarray}
where $Q_s(b_\perp)$ is the commonly defined impact parameter dependent saturation scale. After
carrying out the azimuthal angle integration, one arrives at,
 \begin{eqnarray}
{\cal F}_{x_g}(q_\perp^2, \Delta_\perp^2)\!\!&=&\!\! \int \frac{d^2 b_\perp d^2 r_\perp}{(2\pi)^4}
e^{-iq_\perp\cdot r_\perp-i\Delta_\perp \cdot b_\perp}
 {\exp} \! \left[ -\frac{r_\perp^2 Q_s^2}{4}\right ]
 \label{fd}
  \\
  {\cal F}_{x_g}^{\cal E}(q_\perp^2, \Delta_\perp^2)\!\!&=&\!\!- \!\! \int \frac{d|b_\perp| d|r_\perp|}{(2\pi)^2} J_2(|q_\perp| |r_\perp|) J_2(|\Delta_\perp|
|b_\perp|)  \frac{|r_\perp|^5|b_\perp|^3   }{24} \frac{\partial^2Q_s^2(b_\perp^2)}{\partial^2
b_\perp^2}
 {\exp} \! \left[ -\frac{r_\perp^2 Q_s^2}{4}\right ]
 \label{fde}
\end{eqnarray}
where $J_2$ is the second order Bessel function. These are the main results of our paper.  In the
dilute limit where $q_\perp^2 \gg Q_s^2 \gg \Delta_\perp^2$, one can reproduce Eq.~\ref{dilute}. by
expanding the exponential and keeping the first nontrivial term. This provides us a nice
consistency check. Another observation one can make is that the elliptic gluon distribution
vanishes if color source were uniformly distributed in the transverse plane of a nucleus. Following
the same method, one could also compute the Weizs\"{a}cker-Williams(WW) type gluon elliptic GTMD in
the MV model. However, we leave it for the future study as it is less interesting
phenomenologically~\cite{Hatta:2016dxp}.

 Let us now close this section by presenting a simple numerical estimation.
 It is rather common to  parameterize the $b_\perp$ dependence of the saturation scale as the following
 ~\cite{Kowalski:2003hm,Rezaeian:2012ji},
\begin{eqnarray}
Q_s^2=  Q_{s0}^2 \exp \left[ -\frac{b_\perp^2}{Bg}\right ] {\rm ln }\left [ \frac{1}{r_\perp^2
\Lambda_{QCD}^2}+E \right ]
\end{eqnarray}
We fix the parameters to be: $\Lambda_{QCD}=0.2 \ GeV$, $B_g=8 \ GeV^{-2}$ and $Q_{s0}^2=1 \
GeV^2$. Inserting the above formula into the Eq.[\ref{fde}], the elliptical gluon distribution
takes the form,
\begin{eqnarray}
  {\cal F}_{x_g}^{\cal E}(q_\perp^2, \Delta_\perp^2)=- \int \frac{d|b_\perp| d|r_\perp|}{(2\pi)^2} J_2(|q_\perp| |r_\perp|) J_2(|\Delta_\perp|
|b_\perp|)  \frac{|r_\perp|^5|b_\perp|^3   Q_s^2}{24B_g^2}
 {\exp} \left[ -\frac{r_\perp^2 Q_s^2}{4}\right ]
\end{eqnarray}
We are interested to study how the elliptic gluon distribution is affected by the saturation
effect. To this end, we plot the double ratio defined in the following as the function of
$q_\perp$,
\begin{eqnarray}
{ \cal R}(q_\perp^2, \Delta_\perp^2)=\frac{{\cal F}_{x_g}^{\cal E}/{\cal
F}_{x_g}}{\Delta_\perp^2/q_\perp^2}
\end{eqnarray}
where $\Delta_\perp^2/q_\perp^2$ is the ratio between  the elliptic gluon distribution and the
normal dipole gluon distribution  in the dilute limit. When performing the numerical calculation,
we impose a cut off $7 Gev^{-1}$ for the upper limit of  $|b_\perp|$ integration. This effectively
corresponds to removing the  forward scattering contribution. Our numerical result presented in the
Fig.1 indicates  that the elliptic gluon distribution is suppressed in the saturation regime, while
at high transverse momentum, the double ratio approaches one as expected. We plan to carry out more
detailed numerical analysis in a future publication.
\begin{figure}[t]
\begin{center}
\includegraphics[width=9cm]{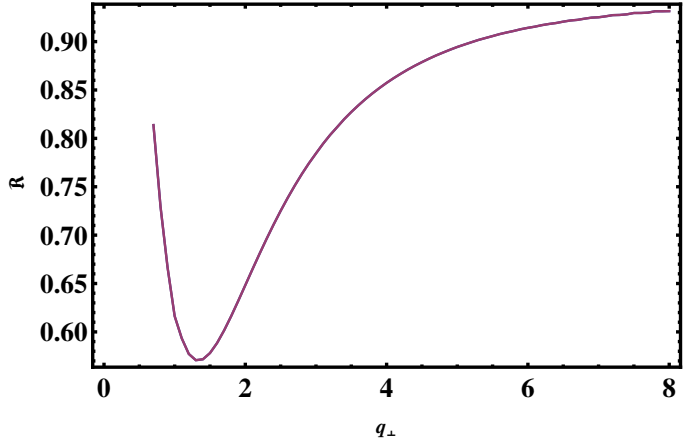}
\caption[] {The double ration ${\cal R}$ as the function of $q_\perp$ for $\Delta_\perp=0.1 \
GeV$.} \label{pomeron}
\end{center}
\end{figure}

\section{Observable}
In Ref.~\cite{Hatta:2016dxp}, it has been proposed to probe the elliptic gluon GTMD by measuring
diffractive dijet production in electron-nucleus  collisions. Such process was first studied in
Ref.~\cite{Altinoluk:2015dpi}. In the back-to-back kinematical limit where the individual jet
transverse momentum $P_\perp$ is much larger than the nucleon recoiled momentum $\Delta_\perp$, a
 $\cos 2( \phi_{P_\perp}-\phi_{\Delta_\perp})$ angular modulation of  the cross section of this
process is sensitive to the elliptic gluon GTMD. Instead of the diffractive dijet production in eA
collisions, we now consider transversely polarized virtual photon-nucleus quasi-elastic scattering
$\gamma^*(q) +A(P)\rightarrow A(P')+X$,  which also offers us the opportunity  to probe the
elliptic gluon distribution via the angular correlation  $\cos
2(\phi_{\epsilon_T^{\gamma^*}}-\phi_{\Delta_\perp})$ where $\epsilon_T^{\gamma^*}$ is the virtual
photon transverse polarization vector.

In the dipole approach, the diffractive cross section for a transversely polarized virtual photon
scattering off nucleus has been computed in Refs.~\cite{Buchmuller:1998jv,Kovchegov:1999kx} and
takes form,
\begin{eqnarray}
\frac{d\sigma_T}{d^2 \Delta_\perp}&=&\pi \int \frac{dz}{z(1-z)}  \int d^2 r_\perp|\Psi^{\gamma^*}_T(Q^2,r_\perp, z)|^2 \nonumber \\
&& \times \left \{ \int \frac{d^2b_\perp}{(2\pi)^2} e^{-ib_\perp \cdot \Delta_\perp} \left [
1-\left \langle \frac{1}{N_c}{\rm Tr} \left [U(b_\perp+\frac{r_\perp}{2})
U^\dag(b_\perp-\frac{r_\perp}{2} )\right ] \right \rangle \right ] \right \}^2
\end{eqnarray}
where $\Psi^{\gamma^*}_T(Q^2,r_\perp, z)$ is the  transversely polarized virtual photon wave
function. The wave function squared can be expressed as,
\begin{eqnarray}
|\Psi^{\gamma^*}_T\!(Q^2,r_\perp, z)|^2\!= 2N_c \sum_q \! \frac{\alpha_{em} e_q^2}{\pi} z(1-z)
\frac{(1-2z)^2(\epsilon_T^{\gamma^*} \!\! \cdot r_\perp)^2+(\epsilon_T^{\gamma^*} \!\! \times
r_\perp)^2}{r_\perp^2} \epsilon_f^2 \left[K_1(|r_\perp|e_f)\right ]^2
\end{eqnarray}
where $e_f^2=Q^2z(1-z)$ and $z$ is the longitudinal momentum fraction of  virtual photon carried by
quark. Here, quark mass is ignored. Inserting the MV model results, the azimuthal independent cross
section is expressed as,
\begin{eqnarray}
\frac{d\sigma_T}{d^2 \Delta_\perp}&=&\pi \int \frac{dz}{z(1-z)} \int  d^2 r_\perp
\Phi(Q^2,r_\perp^2, z)  \left \{ \int \frac{d^2b_\perp}{(2\pi)^2} e^{-ib_\perp \cdot \Delta_\perp}
\left (1- e^{ -\frac{r_\perp^2 Q_s^2}{4}} \right )\right \}^2
\end{eqnarray}
while  the azimuthal dependent cross section reads,
\begin{eqnarray}
\frac{d \sigma_T}{d^2 \Delta_\perp}|_{\cos 2 \phi}&=& \frac{2(\epsilon_T^{\gamma^*} \cdot
\Delta_\perp)^2-\Delta_\perp^2}{24 \Delta_\perp^2} \pi\int \frac{dz}{z(1-z)}
 \int d|r_\perp| \Phi^{\cal E}(Q^2,r_\perp^2, z) |r_\perp|^5
 \nonumber \\
 & \times& \int d|b_\perp|  J_2(|\Delta_\perp|
|b_\perp|)  |b_\perp|^3    \frac{\partial^2Q_s^2(b_\perp^2)}{\partial^2 b_\perp^2}
 e^{ -\frac{r_\perp^2 Q_s^2(b_\perp^2)}{4}}
\int \frac{d^2 b_\perp'}{(2\pi)^2} \left ( 1-e^{ -\frac{r_\perp^2 Q_s^2(b_\perp'^2)}{4}} \right )
\end{eqnarray}
with the scalar part of the wave function being given by,
\begin{eqnarray}
\Phi(Q^2,r_\perp^2, z)&=&2N_c \sum_q \frac{\alpha_{em} e_q^2}{\pi} z(1-z) [z^2+(1-z)^2]
\epsilon_f^2 \left[K_1(|r_\perp|e_f)\right ]^2
\\
\Phi^{\cal E}(Q^2,r_\perp^2, z)&=& 2N_c \sum_q \frac{\alpha_{em} e_q^2}{\pi} z(1-z) [2z^2-2z]
\epsilon_f^2 \left[K_1(|r_\perp|e_f)\right ]^2
\end{eqnarray}
If the virtual photon is induced by a lepton, only the $V_4$
component~\cite{Meng:1991da,Zhou:2008fb} of the corresponding leptonic tensor contributes to the
azimuthal dependent cross section. To be precise, the contribution from this component yields a
$\cos 2 \phi_{e-A}$ azimuthal  modulation where $\phi_{e-A}$ is the angle between the
hadron/nucleus plane and the lepton plane. One notices that the azimuthal asymmetry in this process
offers us a direct access to the second derivative
 of the saturation scale with respect to $b_\perp^2$.  As such, the elliptic gluon GTMD can be easily determined through
 measuring this observable.

\section{Summary}
We derived the elliptic gluon GTMD inside a large nucleus using the MV model. Our result can be
used as the initial condition when implementing small $x$ evolution. We further proposed to probe
the elliptic gluon GTMD through the angular correlation in quasi-elastic virtual photon scattering
off a nucleus. This measurement can, in principle, be carried out at the future EIC. Since a such
study will deepen our understanding how the gluon tomography is  induced by small $x$ dynamics, it
might be worthwhile to pursue a comprehensive numerical analysis of this observable for the typical
EIC accessible kinematical region in a future publication.

\

{\bf Acknowledgments:} I would like to thank D. Mueller for bringing my attention to the small $x$
gluon helicity flip GPDs, and Y. J. Zhou for her help with making the plot. This work has been
supported by the National Science Foundations of China under Grant No. 11160005131608, and by the
Thousand Talents Plan for Young Professionals.

\end {document}